\RequirePackage{fix-cm}
\documentclass[smallcondensed]{svjour3}     
\smartqed  
\usepackage{graphicx}
\usepackage{multirow}
\usepackage{pdfpages}
\usepackage{hyperref}
\journalname{Journal of Intelligent Information Systems}
\begin{document}

\title{LDA-based Term Profiles for Expert Finding in a Political Setting\thanks{All the authors have equally contributed to this paper.}}


\author{Luis M. de Campos \and Juan M. Fern\'andez-Luna \and Juan F. Huete \and Luis Redondo-Exp\'osito}

\authorrunning{de Campos, Fern\'andez-Luna, Huete, Redondo-Exp\'osito} 

\institute{Departamento de Ciencias de la Computaci\'on e Inteligencia Artificial. ETSI Inform\'atica y de Telecomunicaci\'on, CITIC-UGR. Universidad de Granada. Periodista Daniel Saucedo Aranda, s/n, 18014, Granada, Spain\\
\email{\{lci,jhg,jmfluna,luisre\}@decsai.ugr.es}}

\date{Received: date / Accepted: date}

\maketitle

\begin{abstract}
 A common task in many political institutions (i.e. Parliament) is to find politicians who are experts in a particular field. In order to tackle this problem, the first step is to obtain politician profiles which include their interests, and these can be automatically learned from their speeches. As a politician may have various areas of expertise, one alternative is to use a set of subprofiles, each of which covers a different subject. In this study, we propose a novel approach for this task by using latent Dirichlet allocation (LDA) to determine the main underlying topics of each political speech, and to distribute the related terms among the different topic-based subprofiles. With this objective, we propose the use of fifteen distance and similarity measures to automatically determine the optimal number of topics discussed in a document, and to demonstrate that every measure converges into five strategies: Euclidean, Dice, Sorensen, Cosine and Overlap. Our experimental results showed that the scores of the different accuracy metrics of the proposed strategies tended to be higher than those of the baselines for expert recommendation tasks, and that the use of an appropriate number of topics has proved relevant.
\CRclass{10002951.10003317.10003347.10003350  \and \\ 10002951.10003317.10003347.10003354}
\end{abstract}

\section{Introduction}
\label{intro}

The broad context of our work is content-based recommender systems \cite{Bobadilla13} for recommending items to users based on the item's textual description according to user requirements and preferences. When the recommended items are the best people to perform a certain task (a task which is represented as a query to the system), we are referring to the expert finding problem \cite{Balog06}. In our case, the specific context where our recommender system/expert finding methods will be applied is in the political sphere, and the experts are politicians who specialize in certain areas. For example, any national or regional Member of Parliament (MP) sitting on the Agriculture Committee would be expected to have a comprehensive knowledge of the different relevant issues relating to this subject (national and European legislation, problems, initiatives, subsidies, type of crops and their geographical distribution, export-related topics, etc.). The same would happen with other MPs working on other committees such as those concerned with health, culture, the economy, education, etc.

When someone needs to solve a specific problem (e.g. the high temperatures that students have to endure in classrooms towards the end of the academic year) or to find information about a certain issue (e.g. the increase in noise levels at night in a residential area due to the large number of bars), the first hurdle to overcome is to identify who to contact so that a solution can be found. A first line of action might be to use general search engines to find web pages listing politicians but this is time-consuming as the information is not centralized and comes from a variety of different sources, noise is extremely high, and considerable effort is required to obtain relevant information about the correct individuals. A second line of action might be to use a specialized system such as an expert finding system, where textual information from politicians is stored and users can access the information once they have submitted their query. In this instance, the recommender system would return a ranking of relevant MPs instead of documents, thereby enabling the user to contact the relevant MP who could help them resolve their problem.

The textual information about each expert with their areas of interest and expertise is collated from a variety of different sources, for example the reports or documents they have written, or the transcriptions of their interventions in parliamentary debates from which their interests and expertise can be inferred (you are what you speak). The expert finding system would subsequently be fed by a set of textual transcriptions of each of their interventions in the parliamentary plenary or committee sessions.

So that the relevant experts might be found, the areas of interest or expertise of the possible candidates to be recommended must be represented in some way by using a profile \cite{Gauch07}. The most common form of profile associated to a candidate is a set of terms to describe their areas of interest and expertise. When a candidate has a wide range of heterogeneous interests (e.g. a politician is interested in a number of topics such as health, education, and the environment), it may not be ideal to merge them all into a single profile. The reason for this is that certain topics might be underrepresented in relation to others, and it might therefore be difficult to associate these topics with the candidate (if a politician mainly works in the areas of health and education and less so on environmental concerns, their environmental work might go almost unnoticed in their overall profile). Also for the purposes of clarity, it may be interesting to consider a profile as comprising various, more homogeneous subprofiles rather than a single monolithic profile.

For this reason, our objective is to find a way of dividing a heterogeneous single profile, which comprises the terms extracted from all the documents associated with a candidate (in our case, the MP's interventions in parliamentary debates), into a set of homogeneous subprofiles. We believe that if we are able to determine the candidates' areas of interest and expertise more accurately, we will be able to offer better recommendations.
In this paper, we will use latent Dirichlet allocation (LDA) \cite{Blei03,Griffiths04} to try to detect the underlying topics in the document collection. In a previous paper, we tackled the same problem but by using clustering techniques \cite{deCampos19} instead of topic models.

Our use of the LDA technique in this work is quite different from work where LDA or other topic models are combined with expert finding. As we have already mentioned, the most common way of representing documents and profiles (which can be associated to experts) is by using term vectors (bag-of-words representation). Another option is to use a topic model, e.g. LDA to change the problem representation from a representation based on term vectors to another based on topic vectors. Some approaches, therefore, work in the term space, others in the topic space, and even some in both spaces (see the related work). We use LDA to divide the documents into subdocuments which are associated with the different topics but which still remain in the term space rather than the topic space (we essentially use LDA to determine how to distribute the terms among the subprofiles). The subdocuments associated with the same topics (and with the same candidate) are then merged to form the subprofiles. As this process can generate an excessive number of subprofiles for some candidates (some of them having only a small number of terms), we have also developed a general method to reduce the number of subprofiles by selecting only the most relevant topics for each document in a principled way.

The objective of this paper is therefore to determine if LDA is a good method for building expert subprofiles in the term domain in the context of political expert finding. The main contributions of this paper are: (1) we study the use of LDA to generate multiple subprofiles of terms, thematically homogeneous, for expert finding in a political setting; (2) we propose a method to divide documents into homogeneous subdocuments by distributing the appearances of each term in the document, using the matrices generated by LDA from the document collection; (3) we propose a method based on distance and similarity measures to assign to each document an optimal subset of the topics originally associated to this document by LDA; (4) we have carried out an extensive experimentation using a collection of parliamentary documents, comparing our proposals with several alternative models.

This paper is organized as follows: Section \ref{sec:related} examines related work on expert finding and LDA; Section \ref{subprofiles} shows how to divide a document into a set of topic-based subdocuments which will be used to create the user subprofiles; Section \ref{sec:Evaluation} presents the experimental settings in the real context of politician finding; Section \ref{sec:anasize} analyzes the resulting subprofiles; Section \ref{sec:Results} discusses the performance on the expert finding task and compares with the baselines; and, finally, Section \ref{sec:conclusions} outlines our conclusions and presents our future lines of research.

\section{Related work} \label{sec:related}

The aim of expert finding methods (also referred to as expert recommendation, expert search, or expert identification) is to find experts in specific areas. A sign of the increasing interest in these methods is the existence of recent reviews on the subject \cite{AlTaie18,Lin17,Nikzad19,Wang18,Yuan19}.

Expert finding methods have many applications: assigning appropriate reviewers to papers submitted for a conference or journal \cite{Mimno07}, detecting potential answerers in community question answering (CQA) systems \cite{Riahi12,Wang18,Thukral19}, finding collaborators for a project \cite{AlTaie18}, identifying suitable experts \cite{Khuda18} in the academic world \cite{Lin13,Rani15}, social media \cite{Bozzon13,Wei16}, companies, institutions and organizations \cite{Bayati16,Karimzadehgan09} or the whole web \cite{Guan13}. Other potential applications of expert finding are explained in \cite{Nikzad19}. The only expert finding applications in a political domain are those by the authors of this paper \cite{deCampos17a,deCampos17b}.

\subsection{Basic models of expert finding}

The two basic expert finding approaches in terms of representing candidate experts are profile-based methods and document-based methods \cite{Balog12,Momtazi13}. The first method is also called the query-independent approach and the second method is also known as the query-dependent approach \cite{Petkova06}. Profile-based methods build a profile for each candidate expert by concatenating all of the relevant documents. Given a query, these ``macro documents'' (each of which is associated to an expert) can be ranked using document retrieval techniques. This {\em monolithic} approach is also known as the single-document author model \cite{Mimno07}. Document-based methods keep the documents associated to an expert as separate entities and retrieve those documents which are relevant to the query. The candidates are then ranked by combining the relevance scores of the documents associated to each candidate. A specific instance of this method uses the maximum to aggregate the scores of each candidate's documents and is known as the max-document author model \cite{Mimno07}. Different ways of aggregating these scores are studied in \cite{Macdonald08}. Due to the fact that in our case the documents are the interventions associated to each MP, we shall classify this document-based approach as {\em intervention-based}. Although document-based methods are generally considered to be better than profile-based methods \cite{Balog12}, certain studies reached the opposite conclusion \cite{LiuCroft05}. Both monolithic and intervention-based approaches will be used as baselines in our experiments.

In our political domain of interest, relationships between expert candidates (MPs in our case) are not as natural as those appearing between askers and answerers in CQA systems, or between e-mail senders and recipients in other domains. For this reason, we do not explicitly consider expert finding methods based on link analysis \cite{Jurczyk07,Liu17}. Our interest in this review of related work mainly focuses on topic models which are applied to expert finding (and LDA in particular) since our proposal is mainly based on this technique.

\subsection{Expert finding using topic models}

Various approaches are based on probabilistic latent semantic analysis (pLSA) \cite{Qu09,Wu08,Xu12} in the context of CQA. In \cite{Wu08}, pLSA is applied to the collection of questions, users are modeled indirectly, and each user is represented as the average of the topic distributions of the questions in which they are involved. In \cite{Qu09}, users are modeled directly, all the questions relating to a user are grouped into a single document, and pLSA is then applied to this document collection. \cite{Xu12} distinguishes between the two roles that users can play, either as an asker or an answerer. These two roles are jointly modeled using an extension of pLSA.

In a document-based model, the probabilities $p(q|d)$ of query terms given a document are normally estimated using maximum likelihood and Dirichlet smoothing. In \cite{Wu09}, on the other hand, these probabilities $p(q|d)$ are calculated differently using the LDA-learned topics from the document collection as bridges, $p(q|d)=\sum_x p(q|x)p(x|d)$, where $x$ are the different topics. The document-based model is essentially used in \cite{Liu13} in conjunction with a non-uniform prior for the candidates which is built from an LDA modification. In \cite{Liu10}, language models and LDA are combined to predict the best answerers in CQA systems. In both cases, the probabilities of each query term given a user are computed from the user profile comprising all the answers (documents) associated to that user (monolithic profile). In the case of LDA, this computation is performed by averaging over topics, $p(t|userprofile)=\sum_{x} p(t|x)p(x|userprofile)$, where $p(t|x)$ is the distribution of terms given a topic, and $p(x|userprofile)$ is the distribution of topics given the document/userprofile associated to the user, both of which are obtained with LDA. In \cite{Momtazi13}, LDA is applied to the document collection (where the documents include the names of the experts). The fact that the query terms and the experts' names appear, with a high degree of probability, associated to the same topics indicates that the candidate is an expert in the query-associated topics. LDA is used to compute the query term probabilities given the topics and the expert names probabilities given the topics, and these are combined into a formula that is able to compute the probability of a query given an expert. \cite{Zhu11} computes a relevance degree between categories using LDA in CQA systems (whereby categories are represented as topic distributions and relevance is measured with KL divergence). This is combined with a link analysis to find experts in a given target category by considering the information from other relevant categories. \cite{Zhou14} uses LDA on a collection of documents, a single document associated to every user in a CQA system comprising all of their questions, in order to detect similarities between the topics of the askers and answerers. This topic similarity is integrated into a topic-sensitive page rank algorithm in the network that connects askers and answerers.

In \cite{Chi18}, both traditional (static) and dynamic LDA models are considered. In both cases, the topic distributions for the documents associated to the same candidate are averaged. The topic distribution for the query is also computed in order to obtain a similarity measure between the distributions associated to the candidates and to the query. \cite{Daud10} considers temporal topic modeling for expert finding in scientific literature.

In \cite{Mimno07}, the new Author-Persona-Topic (or APT) topic model is proposed to recommend reviewers for submitted papers. In this model, each author can adopt the role of different ``personas'', and these are represented as independent distributions over hidden topics. Each persona represents different topical combinations. Another topic model is the author-topic model (ATM) \cite{Rosen-Zvi04}, an extension of LDA to include authorship by considering relationships between authors, documents, topics, and words. In \cite{Riahi12}, both LDA and another topic model, the Segmented Topic Model (STM), are used in a CQA system. The Segmented Topic Model enables a user profile (comprising in this case the text of questions answered by this candidate) to be modeled as a hierarchical structure, where each question in the profile may have a different distribution over the topics. In \cite{Guo08}, a new topic model is proposed: the User-Question-Answer (UQA) model is an extension of LDA for discovering latent topics contained in terms, categories and users in CQA systems. Users are modeled as pseudo documents by combining all the questions and answers associated to them. In addition to the usual document/user-topic and topic-word distributions, also topic-category distributions are considered. The best performing model considered is a linear combination of the results obtained using topic models with term-based models (a document-based model using BM25).

\cite{Zhao13} proposes the Topic-level Expert Learning (TEL) model and this combines both graph-based link analysis and content-based semantic analysis for expert modeling in CQA. Topic-level Expert Learning is a generative, LDA-based model. Also in the context of community question answering, \cite{Liu15} proposes another combination of LDA (to measure the similarity between candidates) and link analysis. The profile for each candidate is obtained from all of their posted answers. \cite{Yang13} combines topic models and expertise in a probabilistic Topic Expertise Model (TEM) in CQA systems, and this also incorporates information about categories of questions as well as voting information. The CQArank model, which combines user topical interests and expertise learning using TEM with link structure, is then proposed. \cite{Li15} proposes a hybrid model for expert finding in CQA which exploits much of the information available in these systems, such as question tags, votes, and the best answers. These are considered in a link analysis combined with topic analysis, thereby considering both authority and topic similarity between questions and users. In terms of the topic analysis part, the textual content of questions associated to each user is supplemented with the ten most similar words (extracted from Wikipedia using word2vec) for each tag associated to the question. Finally, an extension of LDA, tag-LDA, is used to find the topics associated to each user. In the context of CQA, \cite{Sanpedro14} extends supervised LDA by considering a learning-to-rank paradigm, posing question recommendation as a pairwise ranking problem and taking into account the votes that different experts who have answered a question receive from other users. 

\subsection{Other models of expert finding}

\cite{Serdyukov08} considers the top documents retrieved for the given query. Each of these documents is represented as a mixture of language models of the different experts (in addition to the global language model) because they assume that each document can be associated to several experts. The EM algorithm is then used to estimate the probabilities of query terms given an expert. In \cite{deCampos17a,deCampos17b}, we applied expert finding methods (without using topic models) in the political domain. Basic profile-based (monolithic) and document-based (intervention-based) approaches were used in \cite{deCampos17a}, whereas \cite{deCampos17b} considered subprofiles based on the MPs' involvement on the various parliamentary committees.

In the last few years, and given the success of deep learning models \cite{Alom-etal19}, there is a growing interest in using some of these models for expert finding, particularly doc2vec \cite{Le14} and convolutional neural networks (CNN) \cite{Kim14}. Most of the works are focused on CQA systems, where there are abundant data for training, as in \cite{Dehghan20,Mumtaz19,Wangscichina17}. In \cite{Sohangir18}, doc2vec and CNN are both used to finding experts in financial forums. Some models of this type will be used in our experiments for comparative purposes.

\section{Using LDA to obtain homogeneous subprofiles}
\label{subprofiles}

Let us assume that we have a set of potential candidates for the expert finding problem and a collection $\mathcal{D}$ of documents, each of which is associated with one of the candidates. The interests and expertise of each candidate will be inferred from their associated documents. In our specific case, the candidates will be Members of Parliament (MP) and each document in the collection corresponds to an MP's interventions in parliamentary debates.

As we mentioned previously, our goal is to separate a heterogeneous monolithic profile comprising the terms that appear in every document associated to the candidate into a set of more homogeneous, thematic subprofiles. For this purpose, our first step is to apply LDA to the entire document collection in order to extract the various topics\footnote{We could also apply LDA iteratively only to the documents associated to each candidate, thereby obtaining specific topics for each candidate. In this article, we prefer to explore the usefulness of global topics, although the other approach will be considered in the future.}.

When we apply LDA using $k$ topics\footnote{The number of topics being considered, $k$, is an input parameter of the LDA process.} to a document collection containing $n$ documents and $m$ different terms, we essentially obtain two matrices of order $m\times k$ and $k\times n$, where each entry represents the probabilities $p(t|x)$ (for the probability of relevance of a term $t$ for topic $x$) and $p(x|d)$ (for the probability of topic $x$ being associated to document $d$).

\subsection{Separating documents into homogeneous subdocuments}

Once applied LDA to the document collection, the second step is to separate each document $d$ in the collection into a number of subdocuments $d_i$ (initially $k$ subdocuments) so that each subdocument $d_l$ covers the part of $d$ dealing with the corresponding topic $x_l$. Therefore, for each term $t\in d$, we need a method to determine which subdocument must contain $t$. For example, in a document on the schooling of hospitalized children, LDA would probably identify two main topics: health and education. The terms relating to health (hospital, patient, etc.) should mainly be assigned to one subdocument whereas the terms relating to education (school, teaching materials, etc.) should be assigned to another. This division, however, cannot be exclusive since certain terms may match a number of different topics to a greater or lesser extent (in our example, the term {\it children}). For each term $t$ appearing in $d$, we must therefore decide how many instances of $t$ to assign to each subdocument $d_l$.

Our proposal is to distribute the number of appearances of each term $t$ in document $d$, denoted as $freq(t,d)$, among the subdocuments associated to $d$ proportionally to the probabilities $p(x_l|t,d)$, $l=1,\ldots,k$. It is then necessary to efficiently estimate these probabilities from the results obtained by LDA. In order to do so, we will assume an independence relationship: terms and documents are conditionally independent given the topics, i.e. $p(t|x,d)=p(t|x)$. We use this independence to obtain
$$
p(t,x|d)=p(t|x,d)p(x|d)=p(t|x)p(x|d)
$$
 and also
$$
p(t|d) = \sum_{l=1}^k p(t,x_l|d) = \sum_{l=1}^k p(t|x_l)p(x_l|d)
$$
in order to obtain
\begin{equation}
p(x|t,d)=\frac{p(t,x|d)}{p(t|d)} = \frac{p(t|x)p(x|d)}{\sum_{l=1}^k p(t|x_l)p(x_l|d)}
\label{pxtd}
\end{equation}
In this way, we have a way of computing $p(x|t,d)$ from the LDA matrices.

Once we have the topic probabilities $p(x|t,d)$ for every term $t$ in each document $d$, we distribute the instances of $t$ among the subdocuments of $d$ by computing the product $freq(t,d)*p(x|t,d)$ and rounding up the result to the nearest unit. Because of the rounding up in this process, some instances of the term may be missed or additional instances may even be obtained, thus altering the original count $freq(t,d)$. We detect this situation and, if necessary, we either remove the leftover instances from the subdocuments associated to the least probable topics or add the missing instances to the subdocuments associated to the most probable topics, in order to restore the original count. For example, if the probabilities $p(x|t,d)$ ($k=6$) are $(0.390,0.225,0.157,0.077,0.076,0.075)$ and $freq(t,d)=7$, after multiplying we obtain $(2.730,1.575,1.099,0.539,0.532,0.525)$ and after rounding up we obtain $(3,2,1,1,1,1)$ and we have two extra instances (9 instead of 7). We then remove one instance from each of the two least probable topics, thereby obtaining $(3,2,1,1,0,0)$. If we have the same probabilities but in this case $freq(t,d)=3$, we obtain $(1.170,0.675,0.471,0.231,0.228,0.225)$ and $(1,1,0,0,0,0)$ after multiplying and rounding up, respectively, and we have lost one instance. We then add one instance to the most probable topic, thereby obtaining $(2,1,0,0,0,0)$.
The algorithm which generates the subdocuments associated to a document $d$, using Equation~(\ref{pxtd}) and the rounding method described before (Figure \ref{fig-alg}\footnote{Although other different ways of restoring the original frequency counts are possible we believe that, for practical purposes, the possible differences for the whole process are not important.}).

\begin{figure}[htb]
\centering
\begin{verbatim}
compute p(x|t,d) for each topic x, each document d and each term t in d
for each document d {
   for each term t in d {
      sr=0
      for each topic x {
         s[x]=freq(t,d)*p(x|t,d)
         r[x]=round(s[x])
         sr+=r[x]
      }
      if (sr<freq(t,d))
         add 1 to r[x] for the freq(t,d)-sr topics x having
         greater values of s[x]
      else if (sr>freq(t,d))
              substract 1 from r[x] for the sr-freq(t,d) topics x
              having smaller values of s[x] but having r[x]>0
      for each topic x
         add r[x] instances of term t to the subdocument of d
         associated to topic x
   }
}
\end{verbatim}
\caption{Algorithm to generate subdocuments from the probabilities $p(x|t,d)$}
\label{fig-alg}
\end{figure}

\subsection{Merging subdocuments to obtain homogeneous subprofiles}

After applying the algorithm in Figure \ref{fig-alg} to the document collection, we separate each document $d$ into a set of (at most\footnote{If, for example, the probability $p(x|d)$ of a topic given document $d$ is zero, then the subdocument associated to this topic will be empty.}) $k$ subdocuments, each of which contains those terms of $d$ that are more related to a given topic. The third step is then to use these subdocuments to build the candidates' subprofiles.

Let $\mathcal{D}_i$ be the subset of the $n_i$ documents (not subdocuments) in the original collection $\mathcal{D}$ which are associated to candidate $i$, $\mathcal{D}_i=\{d_{i1},\ldots,d_{in_i}\}$. Each document $d_{ij}\in \mathcal{D}_i$ has been separated into (at most $k$) subdocuments $d_{ij1},\ldots,d_{ijk}$ (in such a way that $d_{ij}$ is the union of all these subdocuments, $d_{ij}=\cup_{l=1}^k d_{ijl}$), where each $d_{ijl}$ is associated to the topic $x_l$. The subprofile of candidate $i$ which is associated to topic $x_l$, $\mathcal{S}(i,l)$, is then built as
\begin{equation}
\mathcal{S}(i,l) = \cup_{j=1}^{n_i} d_{ijl}
\label{subprofile}
\end{equation}
In other words, we concatenate all the terms of the subdocuments corresponding to topic $x_l$ associated to all the documents for candidate $i$.

It is apparent that this process tends to generate a relatively high number of subprofiles for each candidate, although this number may be smaller than $k$ (this would be the case if, for example, $p(x|d)$ were zero for all the documents associated to a candidate for one or several topics). It should be noted that as soon as a non-empty subdocument for a topic $x_l$ has been generated from a document associated to candidate $i$, then this candidate will have a subprofile for topic $x_l$. Some of these subprofiles might be fairly unimportant with only a small number of terms, and these could be problematic when they are used as indexed documents by an information retrieval system. In the following section, we therefore examine various methods to reduce the number of subprofiles by reducing the number of topics associated to each document.

\subsection{Selecting the optimal number of subdocuments: distribution strategies}

The number of topics, $k$, considered may be relatively large depending on the situation\footnote{For example, in the experiments conducted later in this paper, we use $k=24,70,300$.}. Although we do not expect a single document to deal with all these topics, if the probability $p(x|d)>0$ then, strictly speaking, document $d$ will be to some extent about topic $x$. Positive but small probabilities can cause the document $d$ to be separated into many subdocuments, some of which with only a small number of terms associated to them and not particularly informative. For this reason, we propose that a subset of the most probable topics be selected in order to obtain a better division of $d$ into subdocuments.

The problem can be formalized in the following way: given a probability distribution over the $k$ topics, $p=(p_1,p_2,\ldots,p_k)$, with $p_1\geq p_2\geq\ldots\geq p_k$, we want to find the best index, $i_p$, such that the selected topics are $x_1,x_2,\ldots,x_{i_p}$, i.e. the $i_p$ most probable topics. We want to select $i_p$ in some principled way.

It is worth mentioning that there are only $k$ possible solutions, $i_p=1,2,\ldots,k$, which may be associated to  the vectors $I_1=(1,0,0,\ldots,0)$ (selecting only the first, most probable topic), $I_2=(1,1,0,\ldots,0)$ (selecting two topics), $I_3=(1,1,1,0,\ldots,0)$, until $I_k=(1,1,\ldots,1)$ (selecting all the topics).

We can therefore formulate our problem as finding the vector

$$
I_{i_p} = \arg\min_{I_j} Dist(p,I_j)
$$
where $Dist$ is a distance measure between the vectors $p$ and $I_j$. Alternatively,
$$
I_{i_p} = \arg\max_{I_j} Sim(p,I_j)
$$
where $Sim$ is a similarity measure between the vectors $p$ and $I_j$.

We will now study various proposals based on different distance and similarity measures. If we have two k-dimensional vectors $w_1=(w_{11},w_{21},\ldots,w_{k1})$ and $w_2=(w_{12},w_{22},\ldots,w_{k2})$, distance and similarity have been defined in many different ways (see \cite{Cha07} for a survey). Let us begin with the similarity measures.

We can use the Cosine similarity measure

$$
Cos(w_1,w_2)=\frac{\sum_{i=1}^k w_{i1}w_{i2}}{\sqrt{\sum_{i=1}^k w_{i1}^2}\sqrt{\sum_{i=1}^k w_{i2}^2}}
$$
which in our case is
$$
Cos(p,I_j)= \frac{\sum_{i=1}^j p_i}{\sqrt{j}\sqrt{\sum_{i=1}^k p_i^2}}
$$
As the value $\sqrt{\sum_{i=1}^k p_i^2}$ is constant for all $I_j$, the solution in this case is

\begin{equation}
I_{i_p}=\arg\max_{I_j} \frac{\sum_{i=1}^j p_i}{\sqrt{j}}
\label{cosine}
\end{equation}

Another similarity measure is the Dice coefficient and this is defined as

$$
Dic(w_1,w_2)=\frac{2\sum_{i=1}^k w_{i1}w_{i2}}{\sum_{i=1}^k w_{i1}^2+\sum_{i=1}^k w_{i2}^2}
$$

which in our case results in

$$
Dic(p,I_j)= \frac{2\sum_{i=1}^j p_i}{j+\sum_{i=1}^k p_i^2}
$$
The solution in this case is
\begin{equation}
I_{i_p}=\arg\max_{I_j} \frac{\sum_{i=1}^j p_i}{j+\sum_{i=1}^k p_i^2}
\label{dice}
\end{equation}

Alternatively, we can use the Jaccard similarity index and this is defined as

$$
Jac(w_1,w_2)=\frac{\sum_{i=1}^k w_{i1}w_{i2}}{\sum_{i=1}^k w_{i1}^2+\sum_{i=1}^k w_{i2}^2-\sum_{i=1}^k w_{i1}w_{i2}}
$$

In our case, this similarity measure results in

$$
Jac(p,I_j)= \frac{\sum_{i=1}^j p_i}{j+\sum_{i=1}^k p_i^2-\sum_{i=1}^j p_i}
$$

We can prove that although $Dic(p,I_j)$ and $Jac(p,I_j)$ give different values, the ranking generated by these measures on the vectors $I_j$ is always the same and so they generate the same solution.

The Czekanowski similarity measure is defined as:

$$
Cze(w_1,w_2)=\frac{2\sum_{i=1}^k \min(w_{i1},w_{i2})}{\sum_{i=1}^k (w_{i1}+w_{i2})}
$$

which in our case results in

$$
Cze(p,I_j)= \frac{2\sum_{i=1}^j p_i}{j+1}
$$
The solution in this case is:
\begin{equation}
I_{i_p}=\arg\max_{I_j} \frac{\sum_{i=1}^j p_i}{j+1}
\label{sorensen}
\end{equation}

The Ruzicka similarity is defined as

$$
Ruz(w_1,w_2)=\frac{\sum_{i=1}^k \min(w_{i1},w_{i2})}{\sum_{i=1}^k \max(w_{i1},w_{i2})}
$$

In our case, this similarity measure results in

$$
Ruz(p,I_j)= \frac{\sum_{i=1}^j p_i}{j+1-\sum_{i=1}^j p_i}
$$
As in the case of Dice and Jaccard, we can prove that since the ranking generated by $Cze(p,I_j)$ and $Ruz(p,I_j)$ on the vectors $I_j$ is always the same, they generate the same solution.

The last similarity measure that we will consider is the Overlap coefficient
$$
Ove(w_1,w_2)=\frac{\sum_{i=1}^k w_{i1}w_{i2}}{\min(\sum_{i=1}^k w_{i1},\sum_{i=1}^k w_{i2})}
$$

which in our case reduces to

$$
Ove(p,I_j)= \sum_{i=1}^j p_i
$$
which always reaches the maximum at $j=k$. In this case, therefore, the solution is always $I_{i_p}=I_k$ (i.e. using all the topics).

Let us now consider various distance measures. We shall start with the Euclidean distance

$$
Euc(w_1,w_2)=\sqrt{\sum_{i=1}^k (w_{i1}-w_{i2})^2}
$$
In our case, we obtain
$$
Euc(p,I_j)=\sqrt{\sum_{i=1}^j (1-p_i)^2 +\sum_{i=j+1}^k p_i^2} = \sqrt{j+\sum_{i=1}^k p_i^2 -2\sum_{i=1}^j p_i}
$$

We can easily show that $\forall j=1,\ldots,k-1,\, Euc(p,I_j)\leq Euc(p,I_{j+1})$. The best solution is always $I_{i_p}=I_1$ if we use the Euclidean distance when only the most probable topic is used.

Distance can also be measured using the Hamming distance (or the City Block or Manhattan distance):

$$
Ham(w_1,w_2)= \sum_{i=1}^k |w_{i1}-w_{i2}|
$$
In this case, we obtain
$$
Ham(p,I_j)=\sum_{i=1}^j (1-p_i) +\sum_{i=j+1}^k p_i = j - \sum_{i=1}^j p_i + \sum_{i=j+1}^k p_i = j+1 -2\sum_{i=1}^j p_i
$$
As in the previous case, it can be shown that $Ham(p,I_j)\leq Ham(p,I_{j+1})$. Using the Hamming distance, therefore, the best solution is once again $I_{i_p}=I_1$.

The Chebyshev distance is defined as

$$
Che(w_1,w_2)=\max_{i=1}^k |w_{i1}-w_{i2}|
$$

which in our case results in

$$
Che(p,I_j)=\max(\max_{i=1}^j (1-p_i),\max_{i=j+1}^k (p_i)) = \max(1-p_j,p_{j+1}) = 1-p_j
$$
and therefore the minimum distance is always obtained when $j=1$, i.e. once again the best solution is $I_{i_p}=I_1$.

The Sorensen distance is defined as

$$
Sor(w_1,w_2)= \frac{\sum_{i=1}^k|w_{i1}-w_{i2}|}{\sum_{i=1}^k(w_{i1}+w_{i2})}
$$
In our case, this reduces to
$$
Sor(p,I_j)= \frac{j+1 -2\sum_{i=1}^j p_i}{j+1} = 1- \frac{2\sum_{i=1}^j p_i}{j+1}
$$
and this is equal to $1-Cze(p,I_j)$, where $Cze$ is the previously considered Czekanowski similarity. The Sorensen distance, therefore, generates the same solution as Czekanowski.

The Soergel distance is

$$
Soe(w_1,w_2)= \frac{\sum_{i=1}^k|w_{i1}-w_{i2}|}{\sum_{i=1}^k \max(w_{i1},w_{i2})}
$$

In our case, we obtain

$$
Soe(p,I_j)= \frac{j+1 -2\sum_{i=1}^j p_i}{j+1 -\sum_{i=1}^j p_i} = 1-\frac{\sum_{i=1}^j p_i}{j+1-\sum_{i=1}^j p_i}
$$
and this is equal to $1-Ruz(p,I_j)$, where $Ruz$ is the previously studied Ruzicka similarity. The Soergel distance therefore generates the same solution as Ruzicka, which in turn generates the same solution as Czekanowski.

The Kulczynski distance is

$$
Kul(w_1,w_2)= \frac{\sum_{i=1}^k|w_{i1}-w_{i2}|}{\sum_{i=1}^k \min(w_{i1},w_{i2})}
$$

In our case, this is

$$
Kul(p,I_j)= \frac{j+1 -2\sum_{i=1}^j p_i}{\sum_{i=1}^j p_i} = \frac{j+1}{\sum_{i=1}^j p_i}-2
$$

It should be noted that $Kul(p,I_j)=2(1/Cze(p,I_j)-1)$. As this expression is a monotonic decreasing function of the Czekanowski similarity, the Kulczynski distance then generates the same solution as Czekanowski.

The Camberra distance is defined as

$$
Cam(w_1,w_2)= \sum_{i=1}^k \frac{|w_{i1}-w_{i2}|}{w_{i1}+w_{i2}}
$$

In our case, we obtain

$$
Cam(p,I_j)= \sum_{i=1}^j \frac{1-p_i}{1+p_i} + \sum_{i=j+1}^k \frac{p_i}{p_i} = \sum_{i=1}^j \frac{1-p_i}{1+p_i} + k-j
$$

It can easily be proven that $\forall j=1,\ldots,k-1$, $Cam(p,I_j)\geq Cam(p,I_{j+1})$, and so the Camberra distance obtains the minimum at $j=k$ and the solution is always $I_{i_p}=I_k$.

The divergence distance is
$$
Div(w_1,w_2) = 2\sum_{i=1}^k \frac{(w_{i1}-w_{i2})^2}{(w_{i1}+w_{i2})^2}
$$
In our case, the result is

$$
Div(p,Ij)= 2\sum_{i=1}^j \frac{(1-p_i)^2}{(1+p_i)^2} + 2\sum_{i=j+1}^k \frac{p_i^2}{p_i^2} = 2\sum_{i=1}^j \frac{(1-p_i)^2}{(1+p_i)^2} + 2(k-j)
$$

As in the previous case, it can be seen that $\forall j=1,\ldots,k-1,\, Div(p,I_j)\geq Div(p,I_{j+1})$ and therefore the solution is again $I_{i_p}=I_k$.

Finally, the last distance measure that we will consider is the Neyman distance, which is defined as
$$
Ney(w_1,w_2) = \sum_{i=1}^k \frac{(w_{i1}-w_{i2})^2}{w_{i1}}
$$
In our case
$$
Ney(p,I_j)= \sum_{i=1}^j \frac{(1-p_i)^2}{p_i} + \sum_{i=j+1}^k \frac{p_i^2}{p_i} = \sum_{i=1}^j \frac{1}{p_i} +1 -2j
$$

Since it can be shown that $\forall j=1,\ldots,k-1$, $Ney(p,I_j)\leq Ney(p,I_{j+1})$, the solution in this case is always $I_{i_p}=I_1$.

It is worth noting that although we have used many distances and similarities (15 in total), we only obtain five different solutions to our problem. One of these corresponds to the default case of considering all the topics and we shall call this Overlap\footnote{This is obtained by the overlap, Camberra and divergence measures.} ($I_{i_p}=I_k$). Another extreme case is that of the Euclidean distance, which always considers only the most probable topic\footnote{This is also obtained by the Hamming, Chebyshev and Neyman measures.} ($I_{i_p}=I_1$). The other three cases correspond to the Cosine similarity ($I_{i_p}=\arg\max_{I_j} \frac{\sum_{i=1}^j p_i}{\sqrt{j}}$), the Dice similarity\footnote{and also the Jaccard measure.} ($I_{i_p}=\arg\max_{I_j} \frac{\sum_{i=1}^j p_i}{j+\sum_{i=1}^k p_i^2}$), and the Sorensen distance\footnote{This is also obtained by the Czekanowski, Ruzicka, Soergel and Kulczynski measures.} ($I_{i_p}=\arg\max_{I_j} \frac{\sum_{i=1}^j p_i}{j+1}$). We have experimentally determined that Dice is somewhat more selective than Sorensen (i.e. Dice tends to select fewer topics than Sorensen), which in turn is more selective than Cosine. Euclidean is obviously always the most selective method. For example, for $k=5$ let us consider the probability distribution (0.50,0.29,0.19,0.01,0.01). Using the Cosine similarity, we then select the first three topics ($i_p=3$), whereas Sorensen selects the first two topics ($i_p=2$), and Dice (in the same way as Euclidean) only selects the most probable topic ($i_p=1$).

\begin{figure}
\centering
\includegraphics[width=0.95\textwidth]{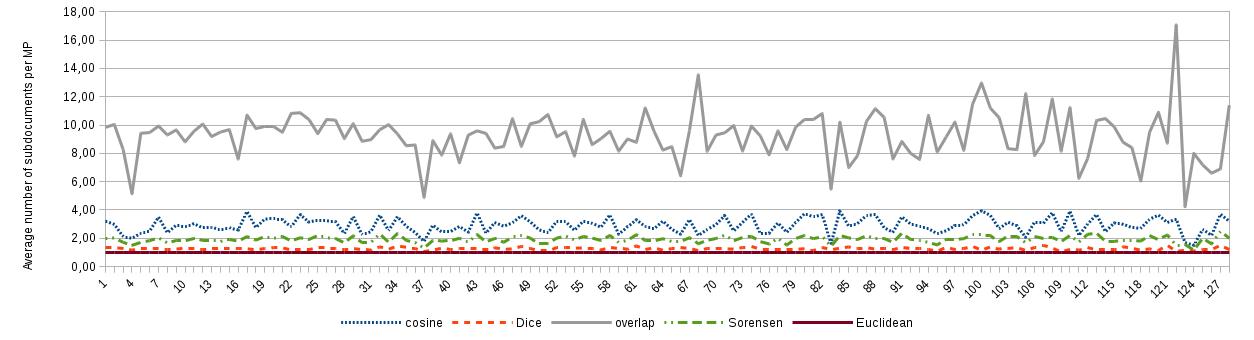}
\caption{Mean number of subdocuments generated from each of the documents associated to each MP, using the five methods ($k=70$)}
\label{numberSubdocs}
\end{figure}

We shall use these five alternatives as distribution strategies for terms in our experiments to test their relative merits. In order to obtain a more precise idea of how these methods perform in our specific political domain, Figure \ref{numberSubdocs} shows the average number of subdocuments generated by each method from each document associated to each MP in our parliamentary document collection (more details about this collection can be found in Section \ref{sec:Evaluation}). Table \ref{mean-max-min} displays the averages of the mean, maximum and minimum number of subdocuments generated from each document for each MP. These data were obtained from an LDA using $k=70$ topics. In the case of the overlap method, the number of subdocuments generated for each document is different from $k=70$ because any topic with a probability $p(x|d)$ of zero is discarded.

\begin{table}[htb]
\caption{Averages of the mean, maximum and minimum number of subdocuments generated from every document associated to each MP ($k=70$)}
\begin{tabular}{llll}
 & mean & max & min \\ \hline
Overlap & 9.25 & 18.66 & 2.39 \\
Cosine & 2.93 & 7.27 & 1.0 \\
Sorensen& 1.92 & 4.16 & 1.0 \\
Dice  & 1.26 & 2.66 & 1.0 \\
Euclidean& 1.0 & 1.0  & 1.0 \\ \hline
\end{tabular}
\label{mean-max-min}
\end{table}

\subsection{Building the optimal number of subdocuments}

The algorithm for generating the subdocuments, displayed in Figure \ref{fig-alg}, must be slightly modified when we apply some of the methods developed in the previous section in order to select a subset of topics. It is important to remember that this selection is based on a probability distribution over the topics. In our context, however, we can use two different distributions, namely $p(x|d)$ (the distribution of topics per document generated by LDA) and $p(x|t,d)$ (the distribution of topics per document and term computed in Equation (\ref{pxtd})), i.e. selecting topics either at the document level or at the document and term level. Since we think that it is preferable to deal with topics at the document level, we will use this option in this paper\footnote{We did in fact conduct experiments with the other option, working at the document and term level, and obtained worse results. The reason for this is probably that terms which are very representative of improbable topics, i.e. those with high $p(x|t,d)$ but low $p(x|d)$, generate their own small subdocuments, resulting in small subprofiles.}.

If we use the distributions $p(x|d)$ to select the optimal number of topics and subdocuments, it is only therefore necessary to replace the original probability $p(x|t,d)$ in Equation (\ref{pxtd}) by $p_o(x|t,d)$:
\begin{equation}
p_o(x|t,d)=\frac{p(t|x)p_n(x|d)}{\sum_{l=1}^k p(t|x_l)p_n(x_l|d)}
\label{poxtd}
\end{equation}
where $p_n(x|d)$ is defined as
\begin{equation}
p_n(x_l|d)= \left\{ \begin{tabular}{cc}
$\frac{p(x_l|d)}{\sum_{j=1}^{i_d} p(x_j|d)}$ & $l=1,\ldots,i_d$ \\
$0$ & $l=i_d+1,\ldots,k$
\end{tabular}
\right.
\label{pnxd}
\end{equation}
where $i_d$ represents the number of the most probable topics selected by the selection method being considered, beginning with the probabilities $p(x|d)$.
In this process, we are simply recomputing the probabilities in Equation (\ref{pxtd}) by using a normalized version, $p_n(x|d)$, of $p(x|d)$ and only considering the selected topics. In this case, we will assume that a document only deals with a small number of topics, and the instances of terms that would originally be assigned to other topics and subsequently eliminated are reassigned to the remaining topics. Each document $d$ is only associated to its $i_d$ most probable topics (according to $p(x|d)$) and each subdocument is associated to exactly one of these topics.

If we combine Equations (\ref{poxtd}) and (\ref{pnxd}), we can write $p_o(x_l|t,d)$ as follows:
\begin{equation}
p_o(x_l|t,d)=\left\{ \begin{tabular}{cc}
$\frac{p(t|x_l)p(x_l|d)}{\sum_{j=1}^{i_d} p(t|x_j)p(x_j|d)}$ & $l=1,\ldots,i_d$ \\
$0$ & $l=i_d+1,\ldots,k$
\end{tabular}
\right.
\label{poxtdbis}
\end{equation}

By proceeding in this way, the subdocuments associated to the selected topics include not only their corresponding terms but also the terms that would be associated to the discarded topics. A different option might be to exclude from the subdocuments the instances of terms associated to discarded topics. In this case, the original document would not be equal to the union of its subdocuments as some terms (or instances of terms) are lost. Previous experiments with this last option led us to discard it mainly because these excluded terms provide some context for the selected subdocuments.

Let us consider the following example: one of the documents contains an intervention where a politician who is a member of the housing committee is speaking about apartments which are allocated to certain social groups, for example young people, groups at risk of social exclusion, or victims of gender violence. The politician's speech contains terms such as {\it woman}, {\it abuse}, etc.  and these terms would correspond to a topic relating to ``Gender Equality''. This intervention could therefore be divided into at least two topics: ``Housing'' and ``Gender Equality''. However, the reference to gender violence is marginal since the main topic of the intervention is housing. After using a method to reduce the number of topics associated to a given document, we would only keep the dominant topic ``Housing'', and therefore the terms {\it woman}, {\it abuse}, etc. would remain associated to this topic. As this politician is a member of the housing committee, then their subprofile is likely to contain many housing-related terms from their interventions that deal primarily with ``Housing'' and relatively few marginal terms (such as {\it woman}). Consequently, if a user submits a query about gender violence, it is reasonable that this politician will not be very relevant (compared with other politicians who deal more specifically with this topic), whereas they might well be erroneously considered more relevant if these words were assigned to a specific, smaller subprofile on ``Gender Equality''. This illustrates the importance of selecting only the most important topics for each document. However, if the query were more specific, e.g. asking about social housing for victims of gender-based violence, then this politician would be more relevant and this would be detected by the system, since both things are combined in the politician's subprofile. In this case, any term which is not specific to a topic but which is included in the corresponding subprofile provides a context so that the other terms can be better understood. If we discard these terms, the system would not be able to highlight the relevance of this politician. In other words, the terms added to a subprofile which are not specific to the corresponding topic probably do not negatively affect system behavior but they can improve it.

\section{Experimental settings} \label{sec:Evaluation}

Our main objective in this paper (and therefore of this evaluation section) is to determine whether the creation of topic-based term subprofiles through the application of LDA is an appropriate method for the expert finding problem. We will therefore consider a political framework in a Parliamentary context.

\subsection{Data sets and evaluation methology}
We will use the Records of Parliamentary Proceedings from the Andalusian Parliament (in Spanish) in its 8th Term of Office\footnote{\url{http://irutai2.ugr.es/ColeccionPA/legislatura8.tgz}}. These records contain all the  speeches in the 5258 initiatives addressed during this term, participating  312  different persons (MPs or stakeholders) in the different sessions, having a total of 12633  interventions. The initiatives will be randomly partitioned into two sets, ensuring that all the initiatives used for testing do not belong to the training set. We randomly sample the 80\% of the initiatives for training purposes, i.e. to run the LDA and build the MPs' subprofiles, and the remaining 20\% is the test set, used to obtain the queries for the expert finding problem. 

\begin{figure}[h]
\centering
\includegraphics[width=\textwidth]{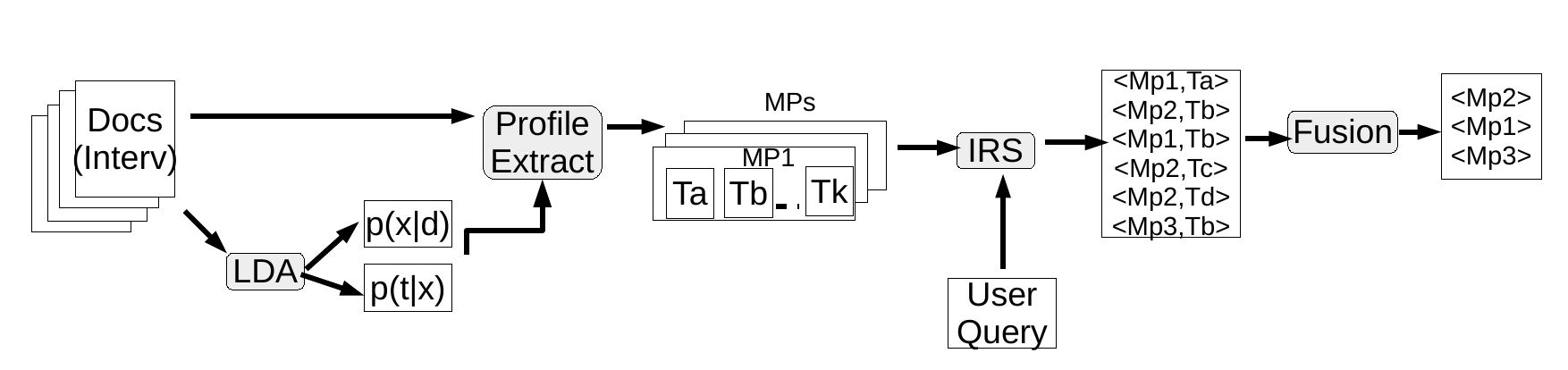}
\caption{Global process\label{fig:proceso}}
\end{figure}

 The random sampling process is then repeated five times to obtain unbiased predictions of our model's performance. The results reported in this paper are the average values of the runs over the five splits or partitions. Three classic IR measures are computed for every query in each test partition: the first two are precision@10 and Normalized Discounted Cumulative Gain (NDCG@10 \cite{Jarvelin02}) and these focus on the top 10 MPs, and the third one is recall@nr and this considers the accuracy of the total number of relevant MPs for each query (nr)\footnote{We do not use recall@10 because in most cases there are more than 10 relevant MPs for each query. It should also be noted that when @nr metrics are considered, the recall and precision values are the same.}. We consider NDCG to be the most appropriate measure of the three for expert finding as we are interested in retrieving the largest number of relevant MPs (as the recall measure would show) but we also wish them to be ranked highest, and this information is given by NDCG.
 
\subsection{LDA implementation}
 The  implementation used for this task was the one presented in the {\em topicmodels} R package. One parameter is the number of topics, $k$, and we had considered three alternatives:
    \begin{itemize}
    \item $k=m*n/t$: classic method \cite{Can90} where, for the problem at hand, $m$ is the number of terms in the Andalusian Parliament collection ($m=4208$), $n$ is the number of MP interventions in the collection ($n=10025$, $80\%$ of the total number of interventions --the training partition), and $t$ is the number of nonzero entries in the document-term matrix ($t=1,702,296$). The value of $k$ is therefore $24$.
    \item $k=\sqrt{n/2}$: another classic method \cite{Kaufman90}, which only considers the number of interventions in the collection. In this case, $k=70$.
    \item $k=300$: this number is higher than those commonly found in LDA-related literature \cite{Chi18,Guo08,Liu10,Liu15,Momtazi13,Wu09,Yang13} although it obtains the best results in \cite{Griffiths04} when compared with both lower and higher values of $k$.
      \end{itemize}
      
      With respect to the other paramenters, and as suggested in \cite{Griffiths04}, the $\alpha$ hyperparameter is set to $50/k$, where $k$ is the number of topics that we wish to extract, and the value of $\beta$ is set at $0.1$. These are the by-default values for this LDA implementation.  Note that the greater the $\alpha$ hyperparameter the greater the possibility that in an initiative (a document) more topics were discussed. The value $\beta$ controls the distribution of the words in the topics, and is fixed for all the experiments\footnote{We have tried other values of $\alpha$ and $\beta$ but the results do not differ too much from the ones presented here, so we decided to include only these by-default values in the paper.}. 
  
\subsection{Training set, profile generation and retrieval system}
If we focus on the training set, Figure \ref{fig:proceso} shows how each input document contains the MP's interventions in an initiative (once their stop words have been removed and stemming performed, and any remaining term that occurs in fewer than $1\%$ of the interventions has been deleted). This set of documents has been used to feed the LDA in order to obtain $p(x|d)$ and $p(t|x)$.

These probabilities are then used to extract the different users' subprofiles, as explained in Section~\ref{subprofiles} using Euclidean, Dice, Sorensen, Cosine and Overlap as the distribution strategies. Each of the obtained subprofiles can be viewed as a document (bag of words) which represents a particular snapshot of the user's interests, ranging from the most heterogeneous one (obtained using the Euclidean measure) to the purest one (obtained by the Overlap measure).

These topic-based subprofiles are then used to feed an IR system. More specifically, given an $MP_i$ each subprofile $<MP_i,T_j>, j=1\ldots k$ will be indexed using the Lucene library\footnote{https://lucene.apache.org/}. The IR system will return the most relevant MPs (the experts) for a given query $q$ submitted by a citizen.

In order to match a query with these subprofiles, we use the Language Model (LM)\footnote{We have also conducted experiments with other models such as BM25 and Vector Space and the results are very similar to those presented in this paper with LM.}. Unfortunately, however, this returns a ranking of topic-subprofiles rather than the list of MPs that we require, and so we need to include a fusion step to generate the final MP ranking. In order to achieve this, we use the $CombLgDCS$ method \cite{deCampos17b}, and this aggregates the scores of all the MP subprofiles while logarithmically devaluing them by taking into account their positions in the ranking in accordance with the following expression:

\begin{equation}
 score(MP_i,q) = \sum_{j} \frac{s(MP_i,T_j)}{
 \log_2(rank(MP_i, T_j) +1)},
 \end{equation}

\noindent where $s(MP_i,T_j)$ denotes the LM score value of the subprofile $<MP_i,T_j>$ regarding a query $q$, and $rank(MP_i,T_j)$ is defined as $1$ if $T_j$ is the first occurrence of a topic for the $MP_i$ in the original ranking and, otherwise, it is the raw value of the position of the topic-based subprofile in this ranking.

\subsection{Query formulation and relevance assessments} \label{sec:queries}
 For each initiative in the test set, only one query will be obtained. This query should be a plausible description of what information the citizen requires and so we have chosen to use its title (a short description of the initiative) and its subjects\footnote{Elements from the EUROVOC thesaurus (https://data.europa.eu/euodp/en/data/dataset/eurovoc) which are manually allocated by Parliament staff to each initiative.}. In this way, we build a short, natural language-based query (in mean 21.6 term occurrences, without any preprocessing) that might be used in a real expert finding task, for example {\em ``Transport concessions for older people. Public Transport, Elderly''}, which is discussed on the ``Equality and Social Welfare Committee''.

Concerning the ground truth used to evaluate our approach, we will only focus on the initiatives in the test set that belong to any of the twenty-six committee sessions (on average a total number of 612 initiatives per partition, i.e. queries, have been used). We can therefore take advantage of our knowledge about the committee in which the issue was discussed and, by extension, we can assume that any MP who is a member of this committee (on average 15.2 MPs per committee\footnote{Only those MPs who have participated at least 10 times in the training set (a total of 132 different MPs) have been considered.}) would be interested in it. Thus, for the above query, only the MPs in the ``Equality and Social Welfare" committee are considered experts for the query. In this way, we form a wide group of relevant experts who could be useful to users.

\section{Analyzing the effects of the distribution strategies}\label{sec:anasize}

The manner in which intervention terms are distributed among the different topics has a direct impact on the learned subprofiles (which will then be indexed) and consequently also on the output ranking of an IR system. In this section, we perform a detailed analysis that will ultimately help us understand the results obtained by our proposal for the expert finding task.

\subsection{Analysis of the subprofile size}

In this section, we will focus on a single partition\footnote{In the other four partitions, the numbers are very similar and tendencies are maintained.} and examine various statistics relating to the subprofile size. For this purpose, Table \ref{tab:Tamanios} shows the total number of subprofiles obtained (row $\#SP$) for each distribution strategy (Euclidean, Dice, Sorensen, Cosine and Overlap) and each $k$, the average value per MP (row Avg. \#SP) and the average number of terms per subprofile (Avg.SP-size).

If we consider the number of topics, $k$, we could state that, as expected, every MP does not deal with every topic, so our approach can recognize some sort of specialization in their speeches. Furthermore, the number of subprofiles increases with $k$ since the LDA is able to identify more specific topics. In this case, however, the subprofiles contain fewer terms and consequently a certain degree of representativeness might be lost.

In terms of the number of subprofiles generated, we can see that Overlap $>$ Cosine $>$ Sorensen $>$ Dice $>$ Euclidean. This is coherent with the way in which the terms are distributed among the topics. As the number of selected topics decreases, we therefore obtain fewer but more heterogeneous subprofiles. The extreme case is the Euclidean metric since all the terms in a document are assigned to a single topic (the most probable topic in the document). On the other hand, the Overlap strategy obtains purer subprofiles since each term occurrence might be assigned to any of the $k$ candidate topics.

As expected, the most probable topic tends to receive more terms, but less probable topics also have also a chance of receiving some terms (generally speaking, a small number of them). In order to illustrate this, we will focus on tiny subprofiles which can be defined as those with fewer than 50 term occurrences (including repetitions). One problem with this type of subprofile is that they might be worthless because of their lack of representativeness and may be viewed as spurious snapshots which are obtained when an MP marginally covers a topic. Table \ref{tab:Tamanios} shows both the absolute number of tiny subprofiles (\#tinySP) and their relative values as a percentage (\#tinySP/\#SP). In this case, we can see that the number of tiny subprofiles obtained might be more troublesome as the number of topics ($k$) or selected topics increases, with the Overlap strategy obviously obtaining large values.

\begin{table}[tbh]

\caption{Analysis of the profile size per distribution technique, where  $m*n/t = 24$ and $\sqrt{n/2} = 70$\label{tab:Tamanios}}
 \resizebox{\textwidth}{!} {
\begin{tabular}{c|lccccc}
\hline
{\bf  $k$} & & {\bf Overlap} & {\bf Cosine} & {\bf Sorensen} & {\bf Dice} & {\bf Euclidean} \\ \hline \hline
 \parbox[t]{2mm}{\multirow{5}{*}{\rotatebox[origin=c]{90}{$m*n/t$}}}
  
  & $\#SP$ & 2987 & 2147 & 1919 & 1664 & 1438 \\
 & Avg. \#SP & 9.73 & 6.99 & 6.25 & 5.42 & 4.68 \\
& Avg. SP-size & 1573.87 & 2190.0 & 2450.48& 2822.10 & 3266.65\\
 & \#tinySP & 559 & 172 & 130 & 98 & 57 \\
  & \%\#tinySP/\#SP & 18.71 & 8.01 & 6.77 & 5.89 & 3.96\\
 \hline
  \parbox[t]{2mm}{\multirow{5}{*}{\rotatebox[origin=c]{90}{$\sqrt{n/2}$}}}
 & $\#SP$ & 10772 & 5774 & 4340 & 3254 & 2762 \\
 & Avg. \#SP  & 35.09 & 18.81 & 14.14 & 10.60 & 9.00 \\
& Avg. SP-size & 442.125 & 819.90 & 1087.50 & 1448.28 & 1704.62 \\
 & \#tinySP & 3696 & 784 & 375 & 196 & 118 \\
  & \%\#tinySP/\#SP & 31.34 &13.58 & 8.64 & 6.02 & 4.27\\
  \hline
  \parbox[t]{2mm}{\multirow{5}{*}{\rotatebox[origin=c]{90}{$300$}}}
 & $\#SP$ & 56122 & 15644 & 7486 & 4182 & 4122 \\
 & Avg. \#SP  & 182.81 & 50.96 & 24.38 & 13.62 & 13.43 \\
 & Avg. SP-size & 164.8&381.75 & 709.52& 1203.13 & 1219.08\\
   & \#tinySP &35986 & 5864 & 1209 & 223 & 225 \\
   & \%\#tinySP/\#SP & 64.12 &37.48 &16.15 & 5.33 & 5.45\\
  \hline
\end{tabular}
}
\end{table}

\subsection{Analysis of the effect on the IR system's output ranking}\label{sec:anaranking}

Once we have analyzed various statistics relating to the number of subprofiles obtained, we shall briefly consider their effect on the IR system's output ranking. It should be noted that in this analysis, we do not consider either the fusion process or the relevance assessments.

Firstly, we will consider an illustrative example: if we analyze the output ranking obtained for the query {\em ``Transport concessions for older people''}, we might expect various subprofiles relating to the topics ``Transportation'' and ``Social Welfare'' to appear in the top results while other topics might be less probable. We are therefore interested in studying the output distribution on topics and it is desirable that this is concentrated in only a small number of them. Although it might also be interesting to study the concentration of MPs in the output ranking, our aim in this case is to discover as many MPs as possible.

In the cases of both topics and MPs, we are talking about the same concept, i.e. the diversity or variability of the obtained results and we therefore need a metric which is able to capture this idea. In probability theory, the classical metric used to measure variability is the entropy of a distribution. Our first step is therefore to use the frequencies of topics/MPs in the ranking to compute both distributions, and then to study how they relate. We would expect to obtain low entropy values over the topic distribution (the queries might only focus on a small number of topics) and at the same time, large entropy values over the MP distribution (the system could find a suitable number of MPs).

\begin{figure}[t]
\centering
\includegraphics[width=0.8\textwidth]{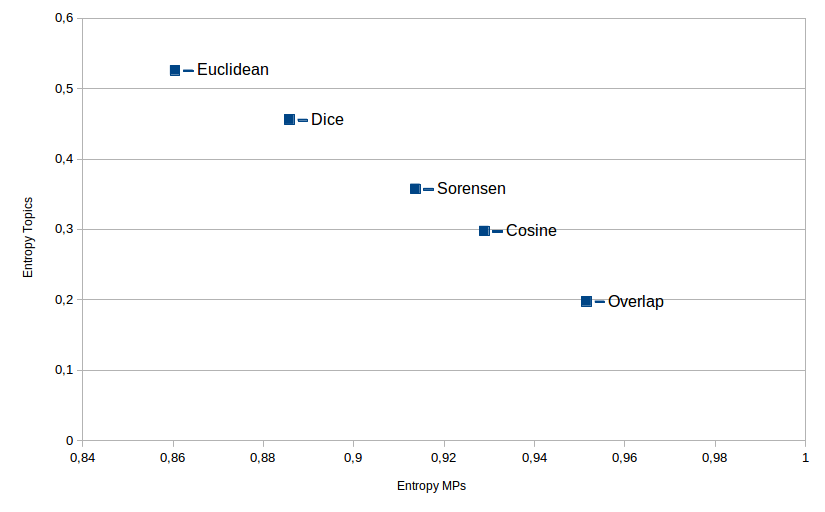}
\caption{Normalized entropy of the distribution of MPs versus the distribution of topics considering $k=\sqrt{n/2}$\label{fig:entropy}}
\end{figure}

The results are shown on the scatterplot in Figure~\ref{fig:entropy} when considering $k=\sqrt{n/2}=70$ and focusing on the top twenty subprofiles $<MP_i,T_j>$\footnote{Similar results are obtained when considering the other values of $k$.}. More specifically, in order to relate both entropies to the distribution strategy, we use the normalized entropy\footnote{The normalized entropy of a distribution over n possible outcomes, with probabilities $p_1,p_2,...,p_n$, is defined as $H_n(p) = -\sum_i \frac{p_i \log_b p_i}{\log_b n}.$}, which gives values between $[0,1]$, so we can use the same scale irrespective of the number of topics and the number of MPs.

From this graph, we can see that the result obtained by the Overlap measure, which obtains the ``purest'' subprofiles, outputs a ranking that focuses on a small number of topics (probably the most representative ones), obtaining also a great diversity on the set of MPs. Although this can be a good option, as we will see later (see Section \ref{sec:Results}), it is not a good choice since the documents in the top positions of the ranking are not representative of MPs' real interests (these subprofiles have a small number of terms and as some of these match the query terms, they therefore obtain high similarity scores). On the other hand, Euclidean strategy, which results in heterogeneous subprofiles (it is not possible to distinguish between intra-document topics), has greater difficulty focusing on a small number of topics in the output ranking. This is because a query term might be located in different topic-based subprofiles, depending on the main topic discussed in the original intervention. Consequently, the output ranking exhibits greater topic variability while returning more subprofiles of the same MP in the top positions (the ranking also somehow represents the heterogeneity of the MPs' speeches). The other strategies present a balance between both situations, but the tendency is clear: increasing subprofile purity (by extracting more topics from a single document) helps to focus on a small number of topics while increasing the MPs' variability.

\section{Performance in the expert finding task} \label{sec:Results}

In this section, we will discuss the quality of the ranking in terms of the accuracy of the different strategies when considering MP recommendation, which is ultimately the main objective of our research. 

\subsection{Baselines}

First, we will consider four baselines, which are the classical approaches found in expert finding literature, and which have already been described in Section \ref{sec:related}: profile-based and document-based methods, but considering both term and topic domains.
\begin{itemize}
\item  Term-based baselines: terms comprise the MPs' profiles and subprofiles, when no LDA is applied
\begin{itemize}
\item Term-based monolithic profiles ($TermMon$): a single profile is built for each MP containing the entire set of terms from all of their interventions.
\item Term-based intervention subprofiles ($TermInt$): there is one subprofile per initiative in which an MP is involved. 
\end{itemize}

In both cases, the MPs' profiles and subprofiles are created according to the training set, and LM is considered as the retrieval model where $CombLgDCS$ is used, when necessary, to obtain the final score for the MPs.

\item  Topic-based baselines: we use LDA to change the representation space from terms to topics, thereby reducing the dimensionality of the problem. A subprofile will therefore be defined through a probability distribution over the set of topics, $p_{i}(x_l), \,l=1,\ldots,k$. These probabilities will be obtained as follows:
  \begin{itemize}
    \item Topic-based monolithic profiles ($TopicMon$): for each $MP_i$, all of their $n_i$ interventions are compiled into a single document, $D_i$, by combining the terms they contain. These compiled documents are used as the input of the LDA algorithm with $k$ topics and the output distribution over topics, $p_i(x_l|D_i)$, is the probability representation of the MP's interests.
    \item Topic-based intervention subprofiles ($TopicInt$): all of the MP's interventions are used as the input of the LDA process by considering $k$ topics. Their associated probability distributions are obtained as the output and these will be used as subprofiles in the topic domain. Each $MP_i$ is therefore represented by a set of probability distributions over the $k$ topics, one for each intervention $d_{ij}$ with $j= 1,\ldots, n_i$, i.e. $\{ p_i(x_l|d_{ij}), \,j=1,\ldots,n_i\}$.
  \end{itemize}

Since the profiles and subprofiles are in the topic domain, we need to transform the queries to the same domain. For this purpose, the LDA output matrices are used to convert the set of queries into topic vectors. Cosine similarity is subsequently used to match query and profiles/subprofiles in this domain. In the case of $TopicInt$, the final score is obtained by applying the $CombLgDCS$ combination method. 
\end{itemize}

In addition to these four baselines, we have also considered for comparison with our approach two other models, coming from the deep learning literature. They are, as we mentioned in Section \ref{sec:related}, doc2vec \cite{Le14} and CNNs \cite{Kim14}.

\begin{itemize}
\item doc2vec is a technique to create document vectors based on shallow neural networks. This is a way of creating an alternative representation of a document collection, as we are doing with the LDA-based topic modelling. After the learning stage, applied to the MPs' interventions, in order to recommend MPs given a query composed of terms, firstly this query is transformed into a query vector using the already learnt doc2vec model. Secondly, the relevance degrees of all document vectors with respect to the query vector are computed by means of the cosine similarity. The final MP ranking is obtained by using again $CombLgDCS$. We have used the Gensim Python implementation of doc2vec, with the default values for some parameters\footnote{Distributed memory as the training algorithm, window of 5 words and a minimal frequency of 5 to consider a word for training.}. We have also selected the best values for the dimensionality of the feature vectors and the number of iterations  for the problem at hand by experimenting with some values. The best results are obtained with a length of 50 and 15 iterations, respectively. 

\item In order to use CNN, we are going to consider this problem as a classification task with as many labels as MPs and using the interventions of each MP as the training set. We used an one-hot-encoding representation with the objective to allow the CNN to compress the input vectors to a dense representation and obtaining in this way our own word embedding that could represent the structure of the language used in the parliament. 
The structure of the CNN is composed by one convolutional layer for 1-dimension data and a kernel size of 10, its respective pooling layer, two hidden dense layers with Rectified Linear Unit (ReLU) as activation function with 5000 and 1000 nodes respectively, and then a final layer with the activation function sigmoid.

Once the CNN has been trained, the obtained values from each of the test documents (queries) represent the strength of each output label, from which we obtain the ranking of MPs. This output works as $N$ networks of one output neuron playing the role of $N$ binary classifiers. Finally, we have to highlight that as loss function we used the keras binary\_crossentropy, keras rmsprop as optimizer and a total of 100 epochs in order to train the CNN\footnote{We tried with higher number of epochs but the accuracy value converges pretty fast in train and validation sets and the accuracy results on test were quite similar.}.
\end{itemize}

\subsection{Results and discussion of the performance at the expert finding task}

The objective of this section is to ascertain whether the application of LDA and the distribution strategies are good approaches for building MP subprofiles, and therefore useful for recommending MPs.

\begin{table}[t]
\caption{Baseline models: Term and topic-based approaches, where  $m*n/t = 24$ and  $\sqrt{n/2} = 70$, doc2vec, CNN and MLP\label{tab:baseline3}}
 \resizebox{\textwidth}{!} {
\begin{tabular}{l|ccc|ccc|ccc}
 \cline{2-10}
 \multicolumn{1}{c}{ } & \multicolumn{3}{|c}{NDCG@10} & \multicolumn{3}{c}{Precision@10} & \multicolumn{3}{c}{Recall@nr} \\
\hline

 TermInt & \multicolumn{3}{c|}{\bfseries{0.7098}} & \multicolumn{3}{c|}{\bf {0.6732}} & \multicolumn{3}{c}{\bf {0.4973}} \\
 TermMon & \multicolumn{3}{c|}{0.6762} &  \multicolumn{3}{c|}{0.6514} & \multicolumn{3}{c}{0.4919} \\ \hline
 doc2vec & \multicolumn{3}{c|}{0.5112} &  \multicolumn{3}{c|}{0.5376} & \multicolumn{3}{c}{0.4130} \\
 CNN     & \multicolumn{3}{c|}{0.1358} &  \multicolumn{3}{c|}{0.1338} & \multicolumn{3}{c}{0.1482} \\ 
 MLP     & \multicolumn{3}{c|}{0.6267} &  \multicolumn{3}{c|}{0.5547} & \multicolumn{3}{c}{0.3357} \\\hline
 \multicolumn{1}{c|}{k } & $m*n/t $ & $\sqrt{n/2}$&  $300$&  $m*n/t $&  $\sqrt{n/2} $ &$300$ &  $m*n/t $&  $\sqrt{n/2} $ &$300$\\
 \hline
 TopicInt & 0.6257 & 0.6045 & 0.6009 & 0.5772 & 0.5581 & 0.6154 & 0.4616 & 0.4393 & 0.4650 \\
 TopicMon & 0.4446 & 0.4636 & 0.5071 & 0.4434 & 0.4583 & 0.5000 & 0.4395 & 0.3840 & 0.3844 \\ \hline
\end{tabular}
}

\end{table}

Firstly, we will analyze the results obtained by the baselines. Table \ref{tab:baseline3} presents the values for the different metrics and the best results are highlighted in bold. From this table, we can conclude that firstly, the term domain performs better than the topic domain (12\% accuracy was lost in the topic domain), and secondly, intervention-based profiles obtain better results than monolithic profiles (this tallies with the results in expert finding literature \cite{Balog12}). Regarding deep learning models, both doc2vec and especially CNNs perform poorly. doc2vec results are only slightly better than those of topic-based monolithic profiles, and always quite worse than the other term-based and topic-based baselines. 

The performance obtained with CNN was not really good. The reason is probably the complexity of this problem (approached in this case as a multilabel classification task with more than 300 classes) for the amount of training data
available (the number of interventions in our parliamentary collection). Deep learning with large models brings benefits when you have a lot of data and a lot of redundancy to be able to generalize, but in specific problems like this with scarce data better solutions can be found \cite{Alom-etal19}. For this reason we also tried to build a simpler model from the CNN approach: we decided to suppress the convolution and pooling layers from the CNN in order to build a Multi-Layer Perceptron (MLP). The obtained results are also displayed in Table \ref{tab:baseline3}. MLP in this case performs much better than CNN, and can be compared for some metrics with topic-based intervention subprofiles, although it is still far from the performance of the best baseline.

We will now discuss the results obtained when considering the different distribution strategies (once the LDA algorithm has been applied). Table \ref{tab:modelos} displays the values obtained by the different metrics, with the best overall value highlighted in bold and the best value for each $k$ written in italics.


From these data, we can see that performance differs slightly between the different metrics and so we will discuss the results by answering the following three research questions:

\begin{table}[t]
\caption{Results for the different distribution strategies (E)uclidean, (D)ice, (S)orensen, (C)osine and (O)verlap, where $m*n/t = 24$ and  $\sqrt{n/2} = 70$\label{tab:modelos}}
\resizebox{\textwidth}{!} {
\begin{tabular}{l|ccc|ccc|ccc}
 \cline{2-10}
\multicolumn{1}{c}{ } & \multicolumn{3}{|c}{NDCG@10} & \multicolumn{3}{c}{Precision@10} & \multicolumn{3}{c}{Recall@nr} \\
\cline{2-10}
 \multicolumn{1}{c|}{ }& $m*n/t $ & $\sqrt{n/2} $&  $300$& $m*n/t$ & $\sqrt{n/2} $&  $300$&  $m*n/t$&  $\sqrt{n/2} $ &$300$ \\
 \hline
E & \textit{0.7372} & 0.7391 & 0.7269 & \textit{0.7131} & 0.7135 & 0.6991 & 0.5349 & 0.5274 & 0.5156 \\
D & 0.7321 & 0.7418 & 0.7268 & 0.7095 & 0.7166 & 0.6998 & \bfseries{\textit{0.5363}} & \textit{0.5295} & \textit{0.5162} \\
S & 0.7273 & \textbf{\textit{0.7482}} & \textit{0.7288} & 0.7052 & \textbf{\textit{0.7197}} & \textit{0.7008} & 0.5321 & 0.5273 & 0.5152 \\
C & 0.7206 & 0.7423 & 0.7245 & 0.6976 & 0.7102 & 0.6943 & 0.5263 & 0.5210 & 0.5056\\
O & 0.6753 & 0.6829 & 0.6512 & 0.6445 & 0.6448 & 0.6116 & 0.4895 & 0.4796 & 0.4613 \\ \hline
\end{tabular}
}

\end{table}

\paragraph{Do the topic distribution strategies have a positive impact on the creation of MP subprofiles?}~

In order to answer this question, we will use the results of term-based baselines, where the terms have not been distributed. In this case, all the strategies (with the exception of Overlap where all the topics are considered) outperform the baselines, independently of the value of $k$. Using the best results, we obtain gains of $5.13\%, 6.46\%$ and $ 7.27\%$ for NDCG@10, precision@10 and recall@nr, respectively, in relation to the best baseline, and the differences are statistically significant (using a t-test) in every case. The reason for the poor performance of Overlap has previously been explained in Section \ref{sec:anaranking} but basically the profiles obtained are not representative of the MPs' interests.

\paragraph{Is the number of topics, $k$, relevant to our task?}~

In this case, it is apparent from Table \ref{tab:modelos} that the use of a large number of topics (e.g. $300$) generally obtains worse results. One explanation for this might be that non-representative subprofiles (some with only a small number of terms) are obtained for some MPs and this can ultimately worsen retrieval effectiveness. When a small number of topics are used, as in $m*n/t = 24$, some of the underlying topics merge. This, in turn, represents a loss in subprofile expressiveness and affects evaluation metric values. In order to address this question, we can conclude that the number of topics matters and must be selected carefully, with $\sqrt{n/2}$ representing a good approach since it performs best in @10 metrics (obtaining statistical significance), whereas for recall@nr $m*n/t$ performs slightly better than $\sqrt{n/2}$ (but in this case there is no statistical significance).

\paragraph{Which strategy should we select?}~

Although it becomes necessary to use a distribution method, it is not clear from the results how this should be selected. In terms of accuracy, and focusing on @10 metrics, we obtain the best result (highlighted in bold-italics) when the Sorensen distribution strategy is used with $k=\sqrt{n/2}$. However, it is also clear from Table \ref{tab:modelos} that there is a dependence in terms of the number of topics considered: when considering a small number of topics, Euclidean seems to be the best alternative whereas Sorensen should be the selected strategy for dealing with a large number of topics. The differences between the two distribution approaches are statistically significant in both cases: the p-values of the t-test are 0.003 and 0.004 for $k=m*n/t$ and $k=\sqrt{n/2}$, respectively. Furthermore, if we focus on recall@nr metric, we can see from the statistic tests that for a fixed $k$, there are no significant differences between Euclidean, Dice and Sorensen. Nevertheless, as we show in Section \ref{sec:anaranking}, there are other criteria that tip the balance in favour of LDA+Sorensen since more homogeneous profiles can be obtained, thereby enabling an adequate level of representativeness of MP interests in addition to good accuracy results.

\section{Conclusions and future research} \label{sec:conclusions}

In this paper, we have shown how the construction of term-based subprofiles by applying the LDA algorithm to mine the underlying topics positively affects the performance of expert recommendation in a political context. We can apply this generative statistical model to the document collection in order to obtain a set of $k$ topics. The probability distributions generated in this process are then used to distribute the terms from the original documents to each topic. We propose five methods which are based on distance and similarity measures to select the most appropriate number of topics for each document. Finally, the chosen number of subprofiles are constructed.

Our experiments clearly show how all of the distribution strategies (with the exception of Overlap) prove to be useful methods which can be used once LDA has been applied to build MP profiles, thereby improving the results of the term-based and topic-based baselines, as well as those of deep learning models. We have also found that the selection of an appropriate number of topics for the output of the LDA algorithm is relevant in the context of this problem, and that $\sqrt{n/2}$ is our preferred option. As we have already mentioned, it is clear that a distribution strategy must be used to generate profiles with a coherent term distribution across the topics, but the selection of the best alternative is not so obvious as this depends on the number of topics considered (a large or small number). Nevertheless, we recommend the Sorensen strategy as it obtains more homogeneous profiles.

In terms of future research, our next step is to determine how the temporal domain could be included when constructing MP subprofiles and whether this would positively affect the general performance of this expert finding problem. In order to perform this task, it might be interesting to use temporal LDA algorithms and this line of research is definitely worth exploring. Another interesting question might be to distribute the document terms among the subprofiles to the level of complete paragraphs rather than single terms in order to better capture the essential topics covered in a document. Finally, the application of our methods to other domains (e.g. academic) is worth pursuing.

\section*{Acknowledgments}
This work has been funded by the Spanish Ministerio de Econom\'ia y Competitividad under projects TIN2016-77902-C3-2-P and PID2019-106758GB-C31, and the European Regional Development Fund (ERDF-FEDER).


\end{document}